%% file: main.tex
\documentclass[11pt]{article}

\usepackage{eepic,epic}
\usepackage{latexsym}
\usepackage{amsfonts}
\usepackage{amsmath}
\usepackage{amssymb}

\makeatletter%
\def\nottoobig#1{{\hbox{$\left#1\vcenter to1.111\ht\strutbox{}\right.\n@space$}}}
\makeatother%

\makeatother%

\makeatletter%

\newcount\hour  \newcount\minutes  \hour=\time  \divide\hour by 60
\minutes=\hour  \multiply\minutes by -60  \advance\minutes by \time
\def\mmmddyyyy{\ifcase\month\or Jan\or Feb\or Mar\or Apr\or May\or Jun\or Jul\or
  Aug\or Sep\or Oct\or Nov\or Dec\fi \space\number\day, \number\year}
\def\hhmm{\ifnum\hour<10 0\fi\number\hour :%
  \ifnum\minutes<10 0\fi\number\minutes}
\def\Draft{{\it Draft of \mmmddyyyy}}

\def\ps@jtsheadings{%
\def\@oddhead{\it\rightmark\hfil\rm\thepage}%
\def\@oddfoot{\hfil\Draft}%
\if@twoside%
\def\@evenhead{\rm\thepage\hfil\it\leftmark}%
\def\@evenfoot{\Draft\hfil}%
\else
\let\@evenhead\@oddhead%
\let\@evenfoot\@oddfoot%
\fi%
}
\def\ps@jtsplain{%
\def\@oddhead{\hfil\Draft}%
\def\@oddfoot{\hfil\rm\thepage\hfil}%
\let\@evenfoot\@oddfoot%
\if@twoside \def\@evenhead{\Draft\hfil} \else \let\@evenhead\@oddhead \fi
}

\def\chaptermark#1{\markboth{\thechapter.\ #1}{\thechapter.\ #1}}%
\def\sectionmark#1{\markright{\thesection.\ #1}}

\def\section{\@startsection {section}{1}{\z@}
    {3.5ex plus1ex minus.2ex}{2.3ex plus.2ex}{\Large\bf}}
\def\subsection{\@startsection{subsection}{2}{\z@}
    {3.25ex plus1ex minus.2ex}{1.5ex plus.2ex}{\large\bf}}
\def\subsubsection{\@startsection{subsubsection}{3}{\z@}
    {3.25ex plus1ex minus.2ex}{1.5ex plus.2ex}{\normalsize\bf}}
\def\paragraph{\@startsection{paragraph}{4}{\z@}
    {3.25ex plus1ex minus.2ex}{1em}{\normalsize\bf}}
\def\subparagraph{\@startsection{subparagraph}{4}{\parindent}
    {3.25ex plus1ex minus.2ex}{1em}{\normalsize\bf}}

\makeatother%

\makeatletter \@beginparpenalty=10000 \makeatother

\def\underl#1 {\leavevmode\let\first=\relax\underli #1 }
\def\underli#1 {\ifx&#1\let\next=\relax\unskip
                \else\let\next=\underli\first\ulinebox{#1}\fi\let\first=\undersp\next}
\def\undersp{\penalty50\ulinebox{\space}\penalty50}
\def\ulinebox#1{\vtop{\hbox{\strut#1}\hrule}}%
\def\unice#1 {\underl #1 & }
\def\desclabel#1{\bf #1\hfil}
\def\desc{\list{}{%
\labelwidth=\leftmargin
\advance \labelwidth by -\labelsep
\let \makelabel=\desclabel}}

\makeatletter %
\newlength{\leftjustindent}
\newlength{\@leftjustindent}
\def\leftjust{\let\\\@leftjustcr\let\end\@endleftjust
  \addtolength{\@leftjustindent}{\leftjustindent}
  \vcenter\bgroup
  \halign\bgroup
    \hbox to\displaywidth{
      \rule{\@leftjustindent}{0ex}$\displaystyle##$\hfill
      }\crcr
}
\def\endleftjust{\crcr\egroup\egroup\endgroup}
\def\@endleftjust#1{\crcr\egroup\egroup\@checkend{#1}\endgroup}
\def\@leftjustcr{\crcr}

\newtheorem{theorem}{Theorem}[section]

\newcommand{\qedblob}{\mbox{\rule[-1.5pt]{5pt}{10.5pt}}}
\def\literalqed{{\ \nolinebreak\hfill\mbox{\qedblob\quad}}}

\def\qed{\literalqed}

\newtheorem{lemma}[theorem]{Lemma}

\newcommand{\singlespacing}{\let\CS=
\@currsize\renewcommand{\baselinestretch}{1}\tiny\CS}
\newcommand{\singlespacingplus}{\let\CS=
\@currsize\renewcommand{\baselinestretch}{1.25}\tiny\CS}
\newcommand{\doublespacing}{\let\CS=
\@currsize\renewcommand{\baselinestretch}{1.75}\tiny\CS}
\newcommand{\draftspacing}{\let\CS=
\@currsize\renewcommand{\baselinestretch}{2.0}\tiny\CS}
\newcommand{\foospacing}{\let\CS=
\@currsize\renewcommand{\baselinestretch}{1.05}\tiny\CS}

\makeatother%

\mathcode`\0="0030      %
\mathcode`\1="0031
\mathcode`\2="0032
\mathcode`\3="0033
\mathcode`\4="0034
\mathcode`\5="0035
\mathcode`\6="0036
\mathcode`\7="0037
\mathcode`\8="0038
\mathcode`\9="0039
\singlespacingplus
\makeatletter
\clubpenalty=\@highpenalty
\widowpenalty=\@highpenalty
\makeatother

\makeatletter
\newcommand{\niceonespacing}{\let\CS=\@currsize\renewcommand{\baselinestretch}{1.1}\tiny\CS}\newcommand{\nicetwospacing}{\let\CS=\@currsize\renewcommand{\baselinestretch}{1.2}\tiny\CS}
\newcommand{\nicethreespacing}{\let\CS=\@currsize\renewcommand{\baselinestretch}{1.3}\tiny\CS}
\newcommand{\singlespacingplusplus}{\let\CS=\@currsize\renewcommand{\baselinestretch}{1.35}\tiny\CS}
\newcommand{\nicefourspacing}{\let\CS=\@currsize\renewcommand{\baselinestretch}{1.4}\tiny\CS}
\newcommand{\nicefivespacing}{\let\CS=\@currsize\renewcommand{\baselinestretch}{1.5}\tiny\CS}
\newcommand{\nicesixpacing}{\let\CS=\@currsize\renewcommand{\baselinestretch}{1.6}\tiny\CS}
\makeatother

\makeatletter
\def\@cite#1#2{[#1\if@tempswa , #2\fi]}
\makeatother

\makeatletter
\def\@citex[#1]#2{\if@filesw\immediate\write\@auxout{\string\citation{#2}}\fi
  \def\@citea{}\@cite{\@for\@citeb:=#2\do
    {\@citea\def\@citea{,\linebreak[0]}\@ifundefined
       {b@\@citeb}{{\bf ?}\@warning
       {Citation `\@citeb' on page \thepage \space undefined}}%
\hbox{\csname b@\@citeb\endcsname}}}{#1}}
\makeatother

\makeatletter
\def\ps@thesis{\def\@oddhead{\hfil\rm\thepage\hfil}\def\@oddfoot{}\def\@evenhead{\hfil\rm\thepage\hfil}\def\@evenfoot{}\def\chaptermark##1{}\def\sectionmark##1{}}
\makeatother

\makeatletter
\def\foobarpt{\textfont\z@\tenrm
  \scriptfont\z@\ninrm \scriptscriptfont\z@\sevrm
\textfont\@ne\tenmi \scriptfont\@ne\ninmi \scriptscriptfont\@ne\sevmi
\textfont\tw@\tensy \scriptfont\tw@\ninsy \scriptscriptfont\tw@\sevsy
\textfont\thr@@\tenex \scriptfont\thr@@\tenex \scriptscriptfont\thr@@\tenex
\def\unboldmath{\everymath{}\everydisplay{}\@nomath\unboldmath
          \textfont\@ne\tenmi
          \textfont\tw@\tensy \textfont\lyfam\tenly
          \@boldfalse}\@boldfalse
\def\boldmath{\@ifundefined{tenmib}{\global\font\tenmib\@mbi\@magscale1\global
        \font\tensyb\@mbsy \@magscale1\global\font
         \tenlyb\@lasyb\@magscale1\relax\@addfontinfo\@xiipt
              {\def\boldmath{\everymath
                {\mit}\everydisplay{\mit}\@prtct\@nomathbold
                \textfont\@ne\tenmib \textfont\tw@\tensyb
                \textfont\lyfam\tenlyb\@prtct\@boldtrue}}}{}\@xiipt\boldmath}%
\def\prm{\fam\z@\tenrm}%
\def\pit{\fam\itfam\tenit}\textfont\itfam\tenit \scriptfont\itfam\ninit
   \scriptscriptfont\itfam\sevit
\def\psl{\fam\slfam\tensl}\textfont\slfam\tensl
     \scriptfont\slfam\tensl \scriptscriptfont\slfam\tensl
\def\pbf{\fam\bffam\tenbf}\textfont\bffam\tenbf
   \scriptfont\bffam\ninbf \scriptscriptfont\bffam\ninbf
\def\ptt{\fam\ttfam\tentt}\textfont\ttfam\tentt
   \scriptfont\ttfam\nintt \scriptscriptfont\ttfam\nintt
\def\psf{\fam\sffam\tensf}\textfont\sffam\tensf
    \scriptfont\sffam\tensf \scriptscriptfont\sffam\tensf
\def\psc{\@getfont\psc\scfam\@xiipt{\@mcsc\@magscale1}}%
\def\ly{\fam\lyfam\tenly}\textfont\lyfam\tenly
   \scriptfont\lyfam\ninly \scriptscriptfont\lyfam\sevly
 \@setstrut \rm}

\makeatother

\newcommand{\scriptnp}{\mbox{\scriptsize\rm NP}}
\newcommand{\np}{\mbox{\rm NP}}
\newcommand{\p}{\mbox{\rm P}}
\newcommand{\sedr}{\mbox{${\cal S}^{\mbox{\scriptsize\rm{}ED}}_{r}$}}
\newcommand{\smdgr}{\mbox{${\cal S}^{\mbox{\scriptsize\rm{}MDG}}_{r}$}}
\newcommand{\parallelnp}{\mbox{$\p_{\|}^{\scriptnp}$}}
\newcommand{\sr}{\mbox{${\cal S}_r$}} 
\newcommand{\vcgeq}{{\tt VC}_{\mbox{\scriptsize ${\tt geq}$}}}
\newcommand{\sigmastar}{\mbox{$\Sigma^\ast$}}
\def\pair#1{{{\langle\!\!~#1~\!\!\rangle}}}

\newcommand{\degree}{\mbox{\it deg}}
\newcommand{\maxdegree}{\mbox{\it max-deg}}
\newcommand{\condition}{\,|\:}
\newcommand\seq{\subseteq}
\newcommand{\mvc}{\mbox{\it mvc}}
\newcommand{\mined}{\mbox{\it min-ed}}
\newcommand{\minmdg}{\mbox{\it min-mdg}}

\newcommand\Lora{\, \Longrightarrow \ }
\newcommand{\sedone}{\mbox{${\cal S}^{\mbox{\scriptsize\rm{}ED}}_{1}$}}
\newcommand{\vcsedone}{{\tt VC}\mbox{-}{\sedone}}
\newcommand{\naturalnumber}{\ensuremath{{  \mathbb{N} }}}
\def\nats{\naturalnumber}
\newcommand{\littlep}{{\rm p}}
\newcommand{\manyone}{\mbox{$\,\leq_{\rm m}^{{\littlep}}$\,}}
\newcommand{\smdgone}{\mbox{${\cal S}^{\mbox{\scriptsize\rm{}MDG}}_{1}$}}
\newcommand{\vcsmdgone}{{\tt VC}\mbox{-}{\smdgone}}
\newcommand{\vc}{{\tt VC}}
\newcommand{\sproof}{\noindent{\bf Proof}\quad}

\foospacing
\title{Recognizing When Heuristics Can Approximate Minimum Vertex Covers Is
  Complete for Parallel Access to NP\footnote{
\protect\singlespacing
This work was
supported in part by the NSF and the DAAD under grant
  NSF-INT-9815095/DAAD-315-PPP-g\"{u}-ab and
by the DFG under grant RO~1202/9-1.
The first author was supported in part by the NSF under grant NSF-CCR-0311021.
The second author was supported in part by a Hei\-sen\-berg Fellowship
of the DFG.
}}

\author{
Edith Hemaspaandra\\
Department of Computer Science \\
Rochester Institute of Technology \\
Rochester, NY 14627, USA\\
{\tt eh@cs.rit.edu}
\and
J\"{o}rg Rothe\\
Institut f\"{u}r Informatik\\
Heinrich-Heine-Universit\"{a}t\\
40225 D\"{u}sseldorf, Germany \\
{\tt rothe@cs.uni-duesseldorf.de}
\and
Holger Spakowski\\
Institut f\"{u}r Informatik\\
Heinrich-Heine-Universit\"{a}t\\
40225 D\"{u}sseldorf, Germany \\
{\tt spakowsk@cs.uni-duesseldorf.de}
}

\date{January 25, 2005}

\makeatletter
\def\@listI{\leftmargin\leftmargini \parsep 4.5pt plus 1pt minus 1pt\topsep
6pt plus 2pt minus 2pt \itemsep  2pt plus 2pt minus 1pt}

\let\@listi\@listI
\@listi
\makeatother

\begin{document}

\typeout{WARNING:  BADNESS used to suppress reporting!  Beware!!}
\hbadness=3000%
\vbadness=10000 %

\pagestyle{empty}
\setcounter{page}{1}

\sloppy

\pagestyle{empty}
\setcounter{footnote}{0}

{\singlespacing

\maketitle

}

\begin{abstract}\noindent
  For both the edge deletion heuristic and the
  maximum-degree greedy heuristic, we study 
  the problem of recognizing those graphs for which
  that heuristic can approximate
  the size of a minimum vertex cover within a constant factor
  of~$r$, where $r$ is a fixed rational number.
  Our main results are
  that these problems are complete for the class of problems solvable via
  parallel access to~$\np$.
  To achieve these main results, we also show that
  the restriction of the vertex cover problem to those graphs for which either
  of these heuristics can find an optimal solution remains NP-hard.


\end{abstract}

\foospacing
\pagestyle{plain}
\sloppy

\section{Introduction}

\noindent
The minimum vertex cover problem is the problem of finding in a given graph a
smallest possible set of vertices that covers at least one vertex of each
edge.  The decision version of the minimum vertex cover problem, ${\tt VC}$,
is one of the standard NP-complete problems~\cite{gar-joh:b:int}.  To cope
with the intractability that appears to be inherent to this problem, various
heuristics for finding minimum vertex covers have been proposed.  Two of the
most prominent such heuristics are the {\em edge deletion heuristic\/} and the
{\em maximum-degree greedy heuristic}, see,
e.g.,~\cite{pap-ste:b:optimization,pap:b:complexity}.  These algorithms run in
linear time and, depending on the structure of the given input graph, may find
a minimum vertex cover, or may provide a good approximation of the optimal
solution.

It is common to evaluate heuristics for optimization problems by analyzing
their worst-case ratio for approximating the optimal solution.  In this
regard, the two heuristics considered behave quite differently: the edge
deletion heuristic always approximates the size of a minimum vertex cover
within a factor of~$2$ and thus achieves the best approximation ratio known,
whereas the maximum-degree greedy heuristic, in the worst case, can have an
approximation ratio as bad as logarithmic in the input size.  The latter
result follows from the early analysis of the approximation behavior of the
greedy algorithm for the minimum set cover problem that was done by
Johnson~\cite{joh:j:approximation},
Lov{\'a}sz~\cite{lov:j:ratio-of-optimal-covers}, and
Chv{\'{a}}tal~\cite{chv:j:greedy-heuristic-for-set-cover} (who studied the
weighted version of minimum set cover).  Note that the vertex cover problem is
the special case of the set cover problem, restricted so that each element
occurs in exactly two sets.  More recently, building on the work of Lund and
Yannakakis~\cite{lun-yan:j:approximating-minimization-problems},
Feige~\cite{fei:j:approximating-set-cover} showed that, unless NP has slightly
superpolynomial-time algorithms, the set cover problem cannot be approximated
within $(1 - \epsilon) \ln n$, where $\epsilon > 0$ and $\ln$ denotes the
natural logarithm.

In this paper, we study the problem of recognizing those input graphs for
which either of the two heuristics can approximate the size of a minimum
vertex cover within a constant factor of~$r$, where $r \geq 1$ is a fixed
rational number.  Let $\sedr$ and $\smdgr$, respectively, denote this
recognition problem for the edge deletion heuristic and for the maximum-degree
greedy heuristic.
Our main results are: 
\begin{description}
\item[Theorem~\ref{thm:sedr-thetatwo-complete}] For each rational number $r$
  with $1 \leq r < 2$, $\sedr$ is $\parallelnp$-complete.
  
\item[Theorem~\ref{thm:smdgr-thetatwo-complete}] For each rational number $r
  \geq 1$, $\smdgr$ is $\parallelnp$-complete.
\end{description}

Here, $\parallelnp$ denotes the class of problems that can be decided in
polynomial time by parallel (i.e., truth-table) access to~$\np$.
Papadimitriou and Zachos~\cite{pap-zac:c:two-remarks} introduced this class
under the name~$\p^{\scriptnp[\mathcal{O}(\log n)]}$, where
``$[\mathcal{O}(\log n)]$'' denotes that at most logarithmically many Turing
queries are made to the $\np$ oracle.  Hemaspaandra~\cite{hem:j:sky}
proved that $\p^{\scriptnp[\mathcal{O}(\log n)]} = \parallelnp$, and
in fact many more characterizations of~$\parallelnp$ are 
known~\cite{koe-sch-wag:j:diff,wag:j:bounded}.  
Other natural $\parallelnp$-complete problems can be found
in the papers by Krentel~\cite{kre:j:optimization},
Wagner~\cite{wag:j:min-max}, and Hemaspaandra et
al.~\cite{hem-hem-rot:j:dodgson,hem-rot:j:max-independent-set-by-greed}.
  
The type of recognition problem studied in this paper was investigated for
other problems and other heuristics as well.  Bodlaender, Thilikos, and
Yamazaki~\cite{bod-thi-yam:j:greedy-for-maximum-independent-sets} defined and
studied the analogous problem for the independent set problem and the
minimum-degree greedy heuristic, which they denoted by~$\sr$.  They proved
that $\sr$ is coNP-hard and belongs to~$\p^{\mbox{\scriptsize $\np$}}$.
Closing the gap between these lower and upper bounds, Hemaspaandra and 
Rothe~\cite{hem-rot:j:max-independent-set-by-greed} proved that $\sr$ is
$\parallelnp$-complete.
As
in~\cite{hem-rot:j:max-independent-set-by-greed}, we obtain
$\parallelnp$-hardness by reducing from a problem (namely, $\vcgeq$, see
Section~\ref{sec:heuristics}) that can be shown to be $\parallelnp$-complete
using the techniques of Wagner~\cite{wag:j:min-max}.  Also, we show that the
vertex cover problem, restricted to those input graphs for which the
heuristics considered can find an optimal solution, remains NP-hard.  We
then lift this NP-hardness lower bound to $\parallelnp$-hardness, which proves
our main results.  This lifting requires a padding technique such that the
given approximation ratio $r$ is precisely met.  In particular, to achieve
$\parallelnp$-hardness of~$\smdgr$ for each rational number $r \geq 1$, we
modify a construction by Papadimitriou and
Steiglitz~\cite{pap-ste:b:optimization} that they use to analyze the
worst-case approximation behavior of the maximum-degree greedy heuristic.

\section{Two Heuristics for the Vertex Cover Problem}
\label{sec:heuristics}

\noindent
We use the following notation.
Fix the two-letter alphabet $\Sigma = \{0,1\}$.  $\sigmastar$ is the
set of all strings over~$\Sigma$. 
Let $\pair{\cdot , \cdot} : 
\sigmastar \times \sigmastar \rightarrow\, \sigmastar$ be a standard
pairing function.
For any set~$L$, let $\| L \|$ denote the number of elements of~$L$.

All graphs considered in this paper are undirected nonempty,
finite graphs without multiple or reflexive edges.  
For any graph~$G$, let $V(G)$ denote the set of vertices of~$G$, and
let $E(G)$ denote the set of edges of~$G$.  
For any vertex $v \in V(G)$, the {\em degree of $v$\/} (denoted by
$\degree_G(v)$) is the number of vertices adjacent to $v$ in~$G$; if $G$ is
clear from the context, we omit the subscript and simply write~$\degree(v)$.
Let $\maxdegree(G) = \max_{v \in V(G)} \degree(v)$ denote the maximum degree
of the vertices of graph~$G$.
Let $G$ and $H$ be two disjoint graphs.  The {\em disjoint union of $G$ and
  $H$\/} is defined to be the graph $U = G \cup H$ with vertex set $V(U) =
V(G) \cup V(H)$ and edge set $E(U) = E(G) \cup E(H)$.  The {\em join of $G$
  and $H$\/} is defined to be the graph $J = G \bowtie H$ with vertex set $V(J)
= V(G) \cup V(H)$ and edge set $E(J) = E(G) \cup E(H) \cup \{\{x,y\}
\condition x \in V(G) \ \wedge\ y \in V(H)\}$.

For any graph~$G$, a subset $C \seq V(G)$ is a {\em vertex cover of
  $G$\/} if for all edges $\{v, w\} \in E(G)$, $\{v, w\} \cap C \neq
\emptyset$.  A vertex cover is said to be a {\em minimum vertex cover of
  $G$\/} if it is of minimum size.  For any graph~$G$, let $\mvc(G)$ denote
the size of a minimum vertex cover of~$G$.  The vertex cover problem (${\tt
  VC}$, for short; see~\cite{gar-joh:b:int}) is defined to be the set of all
pairs $\pair{G,k}$ such that $G$ is a graph, $k$ a positive integer, and
$\mvc(G) \leq k$.

All hardness and completeness results in this paper are with respect to the
polynomial-time many-one reducibility, denoted~$\manyone$.
For sets $A$ and~$B$, we say $A \manyone B$ if and only if there exists a
polynomial-time computable function $f$ such that for all inputs~$x \in
\sigmastar$, $x \in A$ if and only if $f(x) \in B$.

We consider the following two heuristics (see,
e.g.,~\cite{pap-ste:b:optimization,pap:b:complexity}) for finding a minimum
vertex cover of a given graph:
\begin{description}
\item[Edge Deletion Heuristic (ED):] Given a graph~$G$, the algorithm outputs
  a vertex cover $C$ of~$G$.  Initially, $C$ is the empty set.
  Nondeterministically choose an edge $\{u, v\} \in E(G)$, add both $u$ and
  $v$ to~$C$, and delete $u$, $v$, and all edges incident to $u$ and $v$
  from~$G$.  Repeat until there is no edge left in~$G$.
  
\item[Maximum-Degree Greedy Heuristic (MDG):] Given a graph~$G$, the algorithm
  outputs a vertex cover $C$ of~$G$.  Initially, $C$ is the empty set.
  Nondeterministically choose a vertex $v \in V(G)$ of maximum degree, add
  $v$ to~$C$, and delete $v$ and all edges incident to $v$ from~$G$.  Repeat
  until there is no edge left in~$G$.
\end{description}

As mentioned in the introduction, these two heuristics have a quite different
approximation behavior.  While the worst-case ratio of the MDG algorithm is
logarithmic in the input size~\cite{pap:b:complexity,joh:j:approximation}, the
ED algorithm always approximates the optimal solution within a factor of~$2$.
Thus, despite its extreme simplicity, the edge deletion heuristic achieves the
best approximation ratio known for finding minimum vertex
covers~\cite{pap:b:complexity}.

The central question raised in this paper is: How hard is it to determine for
which graphs $G$ either of these two heuristics can approximate the minimum
vertex cover of $G$ within a factor of~$r$, for a given rational number $r
\geq 1$?  Let $\mined(G)$ (respectively, $\minmdg(G)$) denote the minimum size
of the output set of the ED algorithm (respectively, of the MDG algorithm) on
input~$G$, where the minimum is taken over all possible sequences of
nondeterministic choices the algorithms can make.  For any fixed rational $r
\geq 1$, $\sedr$ (respectively, $\smdgr$) is the class of graphs for which ED
(respectively, MDG)
can output a vertex cover of size at most $r$ times the size of a minimum
vertex cover.  Formally,
\begin{eqnarray*}
\sedr & = & \{ G \condition \mbox{$G$ is a graph and $\mined(G)
\leq r\cdot\mvc(G)$} \} ; \\
\smdgr & = & \{ G \condition \mbox{$G$ is a graph and $\minmdg(G)
\leq r\cdot\mvc(G)$} \} .
\end{eqnarray*}

We will prove that for each fixed rational number $r$ with $1 \leq r < 2$,
$\sedr$ is $\parallelnp$-complete, and that for each fixed rational number $r
\geq 1$, $\smdgr$ is $\parallelnp$-complete.  To this end, we give reductions
from the problem $\vcgeq$, which is defined by
\[
\vcgeq = \{ \pair{G,H} \condition \mbox{$G$ and $H$ are graphs such that
  $\mvc(G) \geq \mvc(H)$}\}.
\]
It is known that $\vcgeq$ is $\parallelnp$-complete, cf.
Wagner~\cite{wag:j:min-max}.  A reduction from any problem in $\parallelnp$ to
$\vcgeq$ that in addition has some useful properties (see
Lemma~\ref{lem:vcgeq} below) can easily be obtained using the techniques of
Wagner~\cite{wag:j:min-max}; see~\cite[Thm.~12]{spa-vog:c:thetatwo} 
for an explicit proof of Lemma~\ref{lem:vcgeq}.

\begin{lemma} {\rm{}(cf.~\cite{wag:j:min-max,spa-vog:c:thetatwo})}\quad
\label{lem:vcgeq}
For any set $X \in \parallelnp$, there exists a polynomial-time computable
function $f$ that reduces $X$ to $\vcgeq$ in such a way that for each $x \in
\sigmastar$, $f(x) = \pair{G,H}$ is an instance of $\vcgeq$ and
\begin{eqnarray*}
x \in X     & \Lora & \mvc(G) = \mvc(H) ; \\
x \not\in X & \Lora & \mvc(G) < \mvc(H) .
\end{eqnarray*}
\end{lemma}

\section{The Edge Deletion Heuristic}

\noindent
Lemma~\ref{lem:vcsedone} below states that the vertex cover problem restricted
to graphs in $\sedone$ is NP-hard.  
The reduction $g$ from Lemma~\ref{lem:vcsedone}
will be used in the proof of the main result of this section,
Theorem~\ref{thm:sedr-thetatwo-complete}.  Define the problem
\[
\vcsedone = \{ \pair{G,k} \condition 
\mbox{$G \in \sedone$ and $k \in \nats^+$ and $\mvc(G) \leq k$}\}.
\]

\begin{lemma}  
  \label{lem:vcsedone}
  There is a polynomial-time many-one reduction $g$ from
  $\vc$ to $\vcsedone$ transforming any given graph $G$ into
  a graph $H\in \sedone$ such that 
\begin{equation}
\label{eq:vcsedone}
\mvc(H) = 2 (\mvc(G) + \| V(G) \|).
\end{equation}
  Hence, $\vcsedone$ is $\np$-hard.
\end{lemma}
\sproof
  Given any graph~$G$, we construct the graph $H \in \sedone$ 
as follows.  For each vertex $v\in V(G)$, create a
component $G_v$ that is defined by the vertex set 
$V(G_v) = \{ v_1, v_2, v_3, v_4 \}$ and the edge set
$E(G_v) = \{ \{ v_1, v_2\} , \{ v_3, v_4\}, \{ v_1, v_3\} \}$.

Define the graph $H$ by joining every pair of components that correspond to
adjacent vertices of~$G$:
\begin{eqnarray*}
         V(H) & = & \bigcup_{v\in V(G)} V(G_v); \\
         E(H) & = & \{\{ a_i,b_j\} \condition \mbox{$\{a,b\} \in E(G)$ and 
                      $i,j \in \{ 1, 2, 3, 4 \}$} \}
                    \cup \bigcup_{v\in V(G)} E(G_v).
\end{eqnarray*}  
We now prove Equation (\ref{eq:vcsedone}).
Let $C$ be a minimum vertex cover
of~$G$, i.e., $\mvc(G) = \| C \|$.  Construct a vertex cover $D$ of $H$ as
follows.  For each vertex $v \in C$, add $v_1, v_2, v_3$, and $v_4$ to~$D$;
and for each vertex $w \in V(G) - C$, add $w_1$ and $w_3$ to~$D$.  Hence,
\[
\|D\| = 2 (\| C \| + \| V(G) \|).
\]
Since $mvc(H) \leq \| D \|$, it follows
that $\mvc(H) \leq 2 (\mvc(G) + \| V(G) \|)$.

Conversely, let $D$ be a minimum vertex cover of~$H$, i.e., $\mvc(H) = \|D\|$.
Then, it holds that:
\begin{itemize}
\item for each edge $\{ u,v\}\in E(G)$, $V(G_u) \seq D$ or $V(G_v) \seq D$;
\item for each vertex $v\in V(G)$, $\|D \cap V(G_v)\| \geq 2$.
\end{itemize}
Hence,
\begin{eqnarray*}
\| D \| & \geq & 4 \cdot \mvc(G) + 2 (\|V(G)\| - \mvc(G)) \\
        & = & 2 (\mvc(G) + \|V(G)\|).
\end{eqnarray*}
It follows that $ \mvc(H) \geq 2 (\mvc(G) + \| V(G) \|)$, which proves
Equation~(\ref{eq:vcsedone}).

It remains to prove that $H \in \sedone$.  Let $C$ be a minimum vertex cover
of~$G$.  The edge deletion algorithm can find a vertex cover of $H$ as
follows.  For every vertex $v\in C$, choose the edges $\{ v_1, v_2\}$ and $\{
v_3, v_4\}$.  For the remaining vertices $w \in V(G) - C$, choose the edge $\{
w_1, w_3\}$.  Thus, $\mined(H) = 2 (\mvc(G) + \| V(G) \|)$.  By
Equation~(\ref{eq:vcsedone}), $\mined(H) = \mvc(H)$, so $H \in \sedone$.~\qed

\begin{theorem}  
\label{thm:sedr-thetatwo-complete}
For each rational number $r$ with $1 \leq r < 2$, $\sedr$ is
$\parallelnp$-complete.
\end{theorem}
\sproof
  It is easy to see that $\sedr$ is in $\parallelnp$.
  To prove $\parallelnp$-hardness, let $X$ be an arbitrary set
  in~$\parallelnp$, and let $f$ be the reduction from $X$ to $\vcgeq$ stated
  in Lemma~\ref{lem:vcgeq}.  Fix any rational number $r$ with $1 \leq r
  < 2$, and let $\ell$ and $m$ be integers such that $r = \frac{\ell}{m}$.
  Note that $1 \leq m \leq \ell < 2m$.
  
  For any string $x \in \sigmastar$, let $f(x) = \pair{G_1, G_2}$. 
  Since we can add isolated vertices to any graph $G$ without altering
  $\mvc(G)$, we may without
  loss of generality assume that $\|V(G_1)\| = \|V(G_2)\|$.
  Let $g$ be the reduction from
  Lemma~\ref{lem:vcsedone} that
  transforms any given graph $G$ into a graph $H \in \sedone$ such that
  Equation~(\ref{eq:vcsedone}) holds.
  Let $H_1 = g(G_1)$ and $H_2 =
  g(G_2)$.  Thus, both $H_1$ and $H_2$ are in $\sedone$,
  and for $i \in \{1, 2\}$, we have 
  $\mvc(H_i) = 2 (\mvc(G_i) + \| V(G_i) \|)$. 
  
  We will define a graph $\widehat{H}$ and an integer $k \geq 0$ such that:
\begin{eqnarray}
\label{eq:H1H2-mined}
\mined(\widehat{H}) & = & r (m \cdot \mvc(H_2) + 2km); \\
\label{eq:H1H2-mvc}
\mvc(\widehat{H})   & = & m \cdot \mvc(H_1) + 2km.
\end{eqnarray}

The reduction mapping any given string $x$ (via the pair $\pair{G_1, G_2}$
obtained according to Lemma~\ref{lem:vcgeq} and via the pair $\pair{H_1, H_2}$
obtained according to Lemma~\ref{lem:vcsedone}) to the graph $\widehat{H}$
such that Equations~(\ref{eq:H1H2-mined}) and~(\ref{eq:H1H2-mvc}) are
satisfied will establish that $X \manyone \sedr$.  In particular, from these
equations, we have that:

\begin{itemize}
\item $\mvc(H_2) = \mvc(H_1)$ implies $\mined(\widehat{H}) = r \cdot
  \mvc(\widehat{H})$, and 
\item $\mvc(H_2) > \mvc(H_1)$ implies $\mined(\widehat{H}) > r \cdot
  \mvc(\widehat{H})$.
\end{itemize}
Note that, due to Lemma~\ref{lem:vcgeq}, $\mvc(H_2) \geq \mvc(H_1)$.

\begin{figure}[ht]  
\centering
\input{sedr.eepic}
\caption{The graph $\widehat{H}$ constructed from $H_1$ and~$H_2$.}
\label{fig:sedr}  
\end{figure}
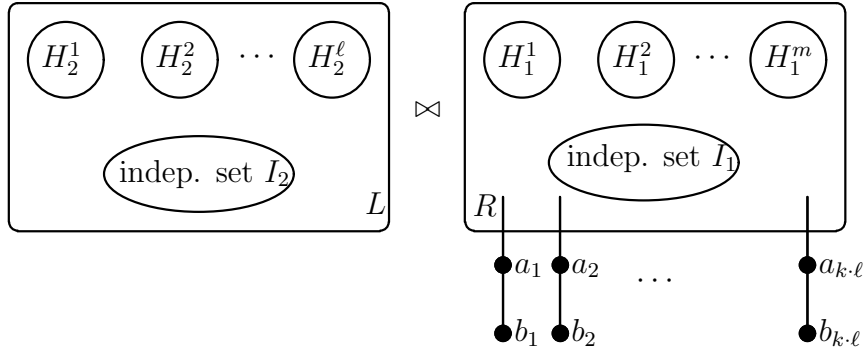

Look at Figure~\ref{fig:sedr} for the construction of $\widehat{H}$ from $H_1$
and~$H_2$.  The graph $\widehat{H}$ consists of two subgraphs, $L$ and~$R$,
that are joined by the join operation, plus some additional vertices and edges
that are connected to~$R$. Formally, let $H_{1}^{1}, H_{1}^{2}, \ldots,
H_{1}^{m}$ be $m$ pairwise disjoint copies of~$H_1$, and let $H_{2}^{1},
H_{2}^{2}, \ldots, H_{2}^{\ell}$ be $\ell$ pairwise disjoint copies
of~$H_2$.  
Let $k = \ell \| V(H_2) \| + m \|V(H_1)\|$.
Let $I_1$ and $I_2$ be independent sets such that $L$ contains
exactly $k(2m-\ell)$ vertices and $R$ exactly $k\ell$
vertices. (This is possible, because
$k(2m-\ell) - \ell \| V(H_2)\|$ is not negative, since 
$2m - \ell \geq 1$, and $k\ell - m \|V(H_1)\|$ is not negative,
since $\ell \geq 1$.)
Let 
$e_i=\{ a_i, b_i\}$ $(1\le i \le k\ell)$ be additional edges. Every vertex
$a_i$ is adjacent to exactly one vertex in $R$, and each vertex in $R$ is
adjacent to exactly one vertex~$a_i$.  The vertices $a_i$ and $b_i$ are not
adjacent to any other vertices. 

\begin{enumerate}
\item
We first determine $\mined(\widehat{H})$.
Let $\widehat{E}$ be a fixed minimum-size output set of the ED algorithm on
input $\widehat{H}$, i.e., $\mined(\widehat{H}) = \| \widehat{E} \|$.
Since $\widehat{E}$ is a vertex cover of $\widehat{H}$, $\widehat{E}$ must
contain $a_i$ or $b_i$ for each $i\in\{ 1,\ldots , k\ell\}$. Since the
ED-algorithm can delete only edges, and $\widehat{E}$ is a minimum-size output
set, it follows that $\widehat{E}$ contains all vertices $a_i$, all vertices
from $R$, and no vertex $b_i$.

Let $C_L$ be a minimum-size output set of the ED-algorithm on input $L$.
By construction of $L$, $\|C_L\| = \ell \cdot \mined(H_2)$.  Thus, since
$H_2 \in \sedone$, $\|C_L\|= \ell\cdot\mvc(H_2)$.

Define $\widehat{E}' = V(R)\cup C_L\cup \bigcup_{i=1}^{k\ell}\{a_i\}$.
It is easy to see that $\widehat{E}'$ is a minimum-size output set of the ED
algorithm on input $\widehat{H}$.
Hence,
\begin{eqnarray*}
   \mined(\widehat{H}) & = & 2k\ell + \ell\cdot\mvc(H_2)\\
                       & = & r(2km + m\cdot\mvc(H_2)).
\end{eqnarray*}
This proves Equation~(\ref{eq:H1H2-mined}).

\item
We now determine $mvc(\widehat{H})$.
Let $\widehat{C}$ be a fixed minimum vertex cover of $\widehat{H}$,
i.e., $\mvc(\widehat{H}) = \|\widehat{C}\|$. Distinguish the following
two cases.
\begin{description}
  \item[Case 1:] $V(R) \seq \widehat{C}$. In this case, $\widehat{C}$ contains all
    vertices from $R$, at least one of $a_i$ or $b_i$ for each $i$, 
    $1\le i \le k\ell$, and a minimum vertex cover of $L$.
    Hence,
    \begin{displaymath}
        \mvc(\widehat{H}) = 2k\ell + \ell\cdot\mvc(H_2).
    \end{displaymath}
  \item[Case 2:] $V(L) \seq \widehat{C}$. In this case, $\widehat{C}$ contains all
    vertices from $L$, each vertex $a_i$, 
    $1\le i \le k\ell$, and a minimum vertex cover of $R$.
    Hence,
    \begin{eqnarray*}
        \mvc(\widehat{H}) & = & k(2m-\ell) + k\ell + m\cdot\mvc(H_1) \\
                          & = & 2km + m\cdot\mvc(H_1).
    \end{eqnarray*}
\end{description}
Since $mvc(H_1) \le \mvc(H_2)$, $m\le \ell$, and $2km\le 2k\ell$, it follows
that
\begin{displaymath}
   mvc(\widehat{H}) = 2km + m\cdot \mvc(H_1).
\end{displaymath}
This proves Equation~(\ref{eq:H1H2-mvc}).
\end{enumerate}
This proves Theorem~\ref{thm:sedr-thetatwo-complete}.~\qed

\section{The Maximum-Degree Greedy Heuristic}

\noindent
Lemma~\ref{lem:vcsmdgone} below states that the vertex cover problem
restricted to graphs in $\smdgone$ is NP-hard.  The proof of
Lemma~\ref{lem:vcsmdgone} 
is reminiscent of a proof by Bodlaender et
al.~\cite[Thm.~4]{bod-thi-yam:j:greedy-for-maximum-independent-sets}, who show
that the independent set problem restricted to graphs for which the
minimum-degree greedy heuristic can find an optimal solution is NP-hard.
The reduction $g$ from Lemma~\ref{lem:vcsmdgone} will be used in
the proof of the main result of this section,
Theorem~\ref{thm:smdgr-thetatwo-complete}.  Define the problem
\[
\vcsmdgone = \{ \pair{G,k} \condition 
\mbox{$G \in \smdgone$ and $k \in \nats^+$ and $\mvc(G) \leq k$}\}.
\]

\begin{lemma}  
\label{lem:vcsmdgone}
  There is a polynomial-time many-one reduction $g$ from
  $\vc$ to $\vcsmdgone$ transforming any given graph $G$ into
  a graph $H\in \smdgone$ such that 
\begin{equation}
\label{eq:vcsmdgone}
\mvc(H) =\mvc(G) + \| E(G) \| (\maxdegree(G) + 1).
\end{equation}
  Hence, $\vcsmdgone$ is $\np$-hard.
\end{lemma}
\sproof
  Given any graph~$G$, we construct the graph $H \in \smdgone$ 
as follows. We replace each edge of $G$ by a gadget that
contains a complete bipartite graph of size $2 (\maxdegree(G) + 1)$.
Formally, $H$ is defined by:
\begin{eqnarray*}
         V(H) & = & V(G) \cup \\
              &   & \bigcup_{e \, = \, \{u,v\} \,\in\, E(G)} 
         \{u_{i}^{e} \condition 1 \leq i \leq \maxdegree(G) + 1\}
         \cup \{v_{i}^{e} \condition 1 \leq i \leq \maxdegree(G) + 1\} ; \\
         E(H) & = & \bigcup_{e \, = \, \{u,v\} \, \in\, E(G)} 
         \left(\{ \{u_{i}^{e}, v_{j}^{e}\} 
         \condition 1 \leq i, j \leq \maxdegree(G) + 1\} 
         \cup \{ \{u, u_{1}^{e}\} \} \cup \{ \{v, v_{1}^{e}\} \}\right).
\end{eqnarray*}  

We now prove Equation~(\ref{eq:vcsmdgone}).  Let $C$ be a minimum vertex cover
of~$G$, i.e., $\mvc(G) = \| C \|$.  Note that $\{u, v\} \cap C \neq \emptyset$
for each edge $\{u, v\}$ in~$E(G)$.  Construct a vertex cover $D$ of $H$ as
follows:
\begin{itemize}
\item $D$ contains all vertices from $C$.
\item For every edge $e = \{ u, v \}$ in $E(G)$, add to~$D$:
\begin{itemize}
\item either all vertices $u_{i}^{e}$, $1 \leq i \leq \maxdegree(G) + 1$, if
  $u \not\in C$ or if both $u$ and $v$ are in~$C$;
\item or all vertices $v_{i}^{e}$, $1 \leq i \leq \maxdegree(G) + 1$, if $v
  \not\in C$.
\end{itemize}
\end{itemize}
It follows that $\mvc(H) \leq \mvc(G) + \| E(G) \| (\maxdegree(G) + 1)$.

Conversely, let $D$ be a minimum vertex cover of~$H$, i.e., $\mvc(H) = \|D\|$.
Construct a vertex cover $C$ of $G$ as follows.  Initially, set $C = D$.  Let
$e = \{ u, v \}$ be any fixed edge in~$E(G)$.  Suppose that at least one
vertex from $\{ u,v\}$ is in~$D$.  Since $D$ is a vertex cover of~$H$, it
contains at least $\maxdegree(G) + 1$ of the vertices $u_{i}^{e}$
and~$v_{i}^{e}$, $1 \leq i \leq \maxdegree(G) + 1$, that correspond to the
edge~$e$.  Remove any $\maxdegree(G) + 1$ such vertices from~$C$.  Suppose now
that neither $u$ nor $v$ is in~$D$.  Since $D$ is a vertex cover of~$H$, it
contains at least $\maxdegree(G) + 2$ of the vertices $u_{i}^{e}$
and~$v_{i}^{e}$, $1 \leq i \leq \maxdegree(G) + 1$, that correspond to the
edge~$e$.  Remove any $\maxdegree(G) + 2$ such vertices from~$C$, and add to
$C$ one of $u$ or $v$ instead.  Since the set $C$ thus obtained is a vertex
cover of~$G$, we have $\mvc(H) \geq \mvc(G) + \|E(G)\| (\maxdegree(G) + 1)$,
which proves Equation~(\ref{eq:vcsmdgone}).

It remains to prove that $H \in \smdgone$.  Let $C$ be a minimum vertex cover
of~$G$.  The maximum-degree greedy algorithm can find a vertex cover of $H$ as
follows.  For every edge $e = \{ u, v \}$ in~$E(G)$, the MDG algorithm on
input $H$ can choose:
\begin{itemize}
\item either all vertices $u_{i}^{e}$, $1 \leq i \leq \maxdegree(G) + 1$, if
  $u \not\in C$ or if both $u$ and $v$ are in~$C$;
\item or all vertices $v_{i}^{e}$, $1 \leq i \leq \maxdegree(G) + 1$, if $v
  \not\in C$.
\end{itemize}
Note that the MDG heuristic can always do so, since every vertex in $V(G)$ has
degree at most $\maxdegree(G)$.  Subsequently, all vertices that are not in
$C$ are isolated. Thus, the MDG algorithm can now choose all vertices
from~$C$.  Hence, $\minmdg(H) = \mvc(G) + \| E(G)\| (\maxdegree(G) + 1)$.  By
Equation~(\ref{eq:vcsmdgone}), $\minmdg(H) = \mvc(H)$, so $H \in
\smdgone$.~\qed

\medskip

Lemma~\ref{lem:mdg-bipartite} below will be used in the proof of
Theorem~\ref{thm:smdgr-thetatwo-complete}.  
The construction of the graph $G$
in this lemma is a modification of a construction given by Papadimitriou and
Steiglitz~\cite[p.~408, Fig.~17-3]{pap-ste:b:optimization}, which shows that
the worst-case approximation ratio of the MDG heuristic can be as bad as
logarithmic in the input size, and so grows unboundedly.  Similar
constructions for achieving the worst-case approximation behavior of the
greedy heuristic solving the more general minimum set cover problem were given
by Johnson~\cite{joh:j:approximation},
Lov{\'a}sz~\cite{lov:j:ratio-of-optimal-covers}, and
Chv{\'{a}}tal~\cite{chv:j:greedy-heuristic-for-set-cover}.

\begin{lemma} 
\label{lem:mdg-bipartite}
For all positive integers $n_1$, $n_2$, $\delta$, and $\mu$ satisfying
\begin{equation}
\label{eq:mdg-bipartite}
\mu (\ln \mu - 2 \ln (\delta + 2) - 1) \geq n_1 + n_2 ,
\end{equation}
there exists a bipartite graph $G$ with the following properties:
\begin{enumerate}
\item \label{eq:mdg-bipartite-1}
$V(G) = V \cup \tilde{V}$ such that $V \cap \tilde{V} = \emptyset$
  and both $V$ and $\tilde{V}$ are independent sets, where
\begin{itemize}
\item $V = \{ u_1, u_2, \ldots , u_{n_1}, w_1, w_2, \ldots , w_{\mu}, z_1,
  z_2, \ldots z_{n_2}\}$ and
\item $\tilde{V} = \{ \tilde{u}_1, \tilde{u}_2, \ldots ,
  \tilde{u}_{n_1}, \tilde{w}_1, \tilde{w}_2, \ldots ,
  \tilde{w}_{\mu} \}$.
\end{itemize}
  
\item \label{eq:mdg-bipartite-2}
$\{ \{ u_i, \tilde{u}_i\} \condition 1 \leq i \leq n_1 \} \cup
  \{ \{ w_i, \tilde{w}_i\} \condition 1 \leq i \leq \mu \} \seq E(G)$.
  
\item \label{eq:mdg-bipartite-3}
  Every vertex~$\tilde{u}_i$, where $1 \leq i \leq n_1$, has degree~$1$.
  
\item \label{eq:mdg-bipartite-4}
  For each induced subgraph $S$ of $G$ that can be obtained by deleting
  vertices from~$V$ such that $V \cap V(S) \neq \emptyset$, it holds that
  $\max_{v \in V \cap V(S)} \degree_S(v) > 
   \max_{v \in \tilde{V}} \degree_S(v) + \delta$.
\end{enumerate}
\end{lemma}
\sproof
  Let the constants $n_1$, $n_2$, $\delta$, and $\mu$ be given such that
  Equation~(\ref{eq:mdg-bipartite}) is satisfied.  We describe the
  construction of the graph~$G$.  As stated in the lemma, the vertex set of
  $G$ is given by $V(G) = V \cup \tilde{V}$, where $V$ and~$\tilde{V}$
  are two disjoint independent sets.
  
  Rename the vertices of $V$ by $V = \{ \alpha_1, \alpha_2, \ldots ,
  \alpha_{n_1 + \mu + n_2}\}$.  Let $\tilde{W} = \{ \tilde{w}_1,
  \tilde{w}_2, \ldots , \tilde{w}_{\mu} \}$.  The edge set of $G$ is
  defined as follows:
\begin{itemize}
\item Create the edges $\{ u_i, \tilde{u}_i\}$ for each $i$ with $1 \leq i
  \leq n_1$ and the edges $\{ w_j, \tilde{w}_j\}$ for each $j$ with $1 \leq
  j \leq \mu$.
  
\item Partition $\tilde{W}$ into $\left\lfloor\frac{\mu}{\delta +
      3}\right\rfloor$ disjoint sets $\tilde{W}_{1}^{\delta + 3},
  \tilde{W}_{2}^{\delta + 3}, \ldots ,
  \tilde{W}_{\left\lfloor\frac{\mu}{\delta + 3}\right\rfloor}^{\delta + 3}$
  of size $\delta + 3$ each, possibly leaving out some vertices
  from~$\tilde{V}$ and taking care that no vertex in
  $\tilde{W}_{i}^{\delta + 3}$ already is connected with~$\alpha_i$, $1
  \leq i \leq \left\lfloor\frac{\mu}{\delta + 3} \right\rfloor$.  For each $i$
  with $1 \leq i \leq \left\lfloor\frac{\mu}{\delta + 3} \right\rfloor$,
  connect $\alpha_i$ with each vertex in $\tilde{W}_{i}^{\delta + 3}$ by an
  edge.
  
\item Partition $\tilde{W}$ into $\left\lfloor\frac{\mu}{\delta + 4}
  \right\rfloor$ disjoint sets $\tilde{W}_{1}^{\delta + 4},
  \tilde{W}_{2}^{\delta + 4}, \ldots ,
  \tilde{W}_{\left\lfloor\frac{\mu}{\delta + 4}\right\rfloor}^{\delta + 4}$
  of size $\delta + 4$ each, possibly leaving out some vertices
  from~$\tilde{V}$ and taking care that no vertex in
  $\tilde{W}_{i}^{\delta + 3}$ already is connected
  with~$\alpha_{\left\lfloor\frac{\mu}{\delta + 3}\right\rfloor + i}$, $1 \leq
  i \leq \left\lfloor\frac{\mu}{\delta + 4} \right\rfloor$.  For each $i$ with
  $1 \leq i \leq \left\lfloor\frac{\mu}{\delta + 4} \right\rfloor$, connect
  $\alpha_{ \left\lfloor\frac{\mu}{\delta + 3}\right\rfloor + i}$ with each
  vertex in $\tilde{W}_{i}^{\delta + 4}$ by an edge.
  
\item Partition $\tilde{W}$ into $\left\lfloor\frac{\mu}{\delta + 5}
  \right\rfloor$ disjoint sets $\tilde{W}_{1}^{\delta + 5},
  \tilde{W}_{2}^{\delta + 5}, \ldots ,
  \tilde{W}_{\left\lfloor\frac{\mu}{\delta + 5}\right\rfloor}^{\delta + 5}$
  of size $\delta + 5$ each, possibly leaving out some vertices
  from~$\tilde{V}$ and taking care that no vertex in
  $\tilde{W}_{i}^{\delta + 3}$ already is connected
  with~$\alpha_{\left\lfloor\frac{\mu}{\delta + 3}\right\rfloor +
    \left\lfloor\frac{\mu}{\delta + 4}\right\rfloor + i}$, $1\leq i \leq
  \left\lfloor\frac{\mu}{\delta + 5}\right\rfloor$.  For each $i$ with $1\leq
  i \leq \left\lfloor\frac{\mu}{\delta + 5}\right\rfloor$, connect $\alpha_{
    \left\lfloor\frac{\mu}{\delta + 3}\right\rfloor +
    \left\lfloor\frac{\mu}{\delta + 4}\right\rfloor + i}$ with each vertex in
  $\tilde{W}_{i}^{\delta + 5}$ by an edge.
  
\item Continue in this way until all vertices $\alpha_i$ are connected with
  vertices in~$\tilde{W}$.
\end{itemize}

The construction is possible, since Equation~(\ref{eq:mdg-bipartite}) implies
\begin{equation}
       \left\lfloor\frac{\mu}{\delta + 3}\right\rfloor + 
       \left\lfloor\frac{\mu}{\delta + 4}\right\rfloor + \cdots + 
       \left\lfloor\frac{\mu}{\mu - 1}\right\rfloor \geq
       n_1 + \mu + n_2 ,
\end{equation}
and thus there are enough vertices in~$\tilde{W}$.  To see why, note that 
\begin{eqnarray}
\lefteqn{ \left\lfloor\frac{\mu}{\delta + 3}\right\rfloor + 
       \left\lfloor\frac{\mu}{\delta + 4}\right\rfloor + \cdots + 
       \left\lfloor\frac{\mu}{\mu - 1}\right\rfloor } \nonumber \\
 &  =   & \left\lfloor\frac{\mu}{1}\right\rfloor + 
       \left\lfloor\frac{\mu}{2}\right\rfloor + \cdots + 
       \left\lfloor\frac{\mu}{\mu}\right\rfloor 
 - \left( \left\lfloor\frac{\mu}{1}\right\rfloor + 
       \left\lfloor\frac{\mu}{2}\right\rfloor + \cdots + 
       \left\lfloor\frac{\mu}{\delta + 2}\right\rfloor \right) - 1 \nonumber \\
\label{eq:mdg-bipartite-imply-1}
 & \geq & \mu \ln \mu - \mu 
\left( \frac{1}{1} + \frac{1}{2} + \cdots + \frac{1}{\delta + 2}\right) \\
\label{eq:mdg-bipartite-imply-2}
 & \geq & \mu \ln \mu - 2 \mu \ln(\delta + 2) .
\end{eqnarray}
Equations~(\ref{eq:mdg-bipartite-imply-1})
and~(\ref{eq:mdg-bipartite-imply-2}) hold, since $\frac{1}{2} + \frac{1}{3} +
\cdots + \frac{1}{n} \leq \int\limits_{1}^{n} \frac{1}{x} \, dx = \ln n - \ln
1 = \ln n$ implies for large enough~$n$:
\begin{eqnarray*}
\frac{1}{1} + \frac{1}{2} + \cdots + \frac{1}{n} \leq 2 \ln n 
& \mbox{\quad and \quad} &
\left\lfloor\frac{n}{1}\right\rfloor + 
\left\lfloor\frac{n}{2}\right\rfloor + \cdots + 
\left\lfloor\frac{n}{n}\right\rfloor \geq 1 + n \ln n .
\end{eqnarray*}

It is evident from the construction that $G$ has all required properties.  In
particular, to see why Property~\ref{eq:mdg-bipartite-4} holds, let $S$ be any
induced subgraph of $G$ that can be obtained by deleting vertices from~$V$
such that $V \cap V(S) \neq \emptyset$.  Let $y_S = \max_{v \in V \cap V(S)}
\degree_S(v)$.  By construction, $S$ can have only edges of the form $\{u_i,
\tilde{u}_i\}$ or $\{w_j, \tilde{w}_j\}$ or edges that are added during
the stages $\delta + 3, \delta + 4, \ldots , y_S$, where $\delta + i$ denotes
the stage in which $\tilde{W}$ is partitioned into subsets of size $\delta +
i$.  It follows that
\[
\max_{v \in \tilde{V}} \degree_S(v)
\leq 1 + y_S - (\delta + 3) + 1 
= y_S - \delta - 1
< y_S - \delta,
\]
which proves the lemma.~\qed
\begin{theorem}  
\label{thm:smdgr-thetatwo-complete}
For each rational number $r \geq 1$, $\smdgr$ is $\parallelnp$-complete.
\end{theorem}

\sproof 
  It is easy to see that $\smdgr$ is in $\parallelnp$.  To prove
  $\parallelnp$-hardness of $\smdgr$, let $X$ be an arbitrary set
  in~$\parallelnp$, and let $f$ be the reduction from $X$ to $\vcgeq$ stated
  in Lemma~\ref{lem:vcgeq}.  For any string $x \in \sigmastar$, let $f(x) =
  \pair{G_1, G_2}$.

  It is convenient to consider the special case of $r=1$ and the case of $r>1$
  separately in the proof of Theorem~\ref{thm:smdgr-thetatwo-complete}.  
We start by proving that $\smdgone$ is $\parallelnp$-complete.  
We will define a graph
$\widehat{G}$ and an integer $q \geq 0$ such that:
\begin{eqnarray}
\label{eq:minmdg-special}
\minmdg(\widehat{G}) & = & \mvc(G_2) + q ; \\
\label{eq:mvcmdg-special}
\mvc(\widehat{G})    & = & \mvc(G_1) + q .
\end{eqnarray}

The reduction mapping any given string $x$ (via the pair $\pair{G_1, G_2}$
obtained according to Lemma~\ref{lem:vcgeq}) 
to the graph $\widehat{G}$ such that
Equations~(\ref{eq:minmdg-special}) and~(\ref{eq:mvcmdg-special}) are
satisfied will establish that $X \manyone \smdgone$.  In particular, from
these equations, we have that:
\begin{itemize}
\item $\mvc(G_2) = \mvc(G_1)$ implies $\minmdg(\widehat{G}) =
  \mvc(\widehat{G})$, and
\item $\mvc(G_2) > \mvc(G_1)$ implies $\minmdg(\widehat{G}) >
  \mvc(\widehat{G})$.
\end{itemize}
Note that, due to Lemma~\ref{lem:vcgeq}, $\mvc(G_2) \geq \mvc(G_1)$.
   
We now describe the construction of~$\widehat{G}$.  
Let $g$ be the reduction from Lemma~\ref{lem:vcsmdgone} and let $H_2 = g(G_2)$.
Thus, $H_2$ is in $\smdgone$ and, by Equation~(\ref{eq:vcsmdgone}),
\begin{equation}
\label{eq:vcsmdgone-2}
\mvc(H_2) = \mvc(G_2) + \| E(G_2) \| (\maxdegree(G_2) + 1).
\end{equation}
Since one can add isolated vertices to any graph $G$
without affecting the values of $\mvc(G)$ or $\minmdg(G)$, we may without loss
of generality assume that
\begin{equation}  
\label{eq:size-of-H2}
        \|V(H_2)\| =  \|V(G_1)\| + \|E(G_2)\| (\maxdegree(G_2) + 1).
\end{equation}

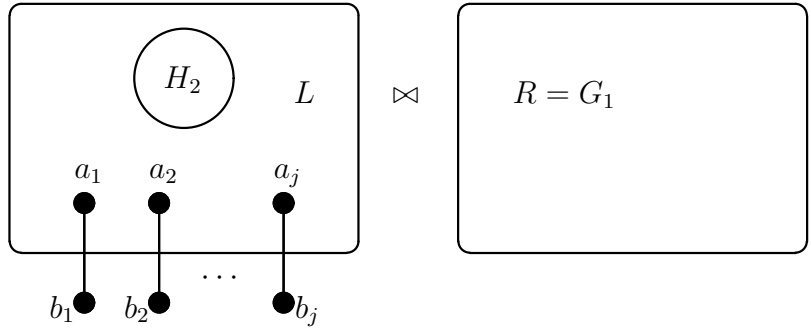
\begin{figure}[ht]  
\centering
\input{smdgone.eepic}
\caption{The graph $\widehat{G}$ constructed from $G_1$ and~$H_2$.}
\label{fig:smdgone}  
\end{figure}

Look at Figure~\ref{fig:smdgone} for the construction of~$\widehat{G}$ 
from $G_1$ and~$H_2$.  The graph $\widehat{G}$ consists of two subgraphs, 
$L$ and~$R$, that
are joined by the join operation, plus some additional vertices and edges that
are connected to~$L$.  Formally, choose $2j$ new vertices $a_i$ and~$b_i$, $1
\leq i \leq j$, where $j$ is a fixed integer large enough such that the degree
of each vertex in $R$ is larger than the maximum degree of the vertices
in~$L$.  Note that the degree of each vertex in $R$ must remain larger
than the degree of any vertex in $L$ even after some vertices have been
removed from~$R$.

Let $B$ be the bipartite matching with the vertex set 
\[
V(B) = \{ a_i \condition 1 \leq i \leq j \} \cup \{ b_i \condition 1 \leq i
\leq j \}
\] 
and the edge set $E(B) = \{\{a_i, b_i\} \condition 1 \leq i \leq j \}$.
Let $R =
G_1$, and let $L$ be the graph with the vertex set $V(L) = \{ a_i \condition 1
\leq i \leq j \} \cup V(H_2)$ and the edge set $E(L) = E(H_2)$.  The graph
$\widehat{G}$ is defined by forming the join $L \bowtie R$, i.e., there are 
edges
connecting each vertex of $L$ with each vertex of~$R$, plus attaching the
vertices~$b_i$, $1 \leq i \leq j$, to $L$ by adding the $j$ edges from~$E(B)$.
    
We first consider $\minmdg(\widehat{G})$.  By our choice of~$j$, each vertex in
$R$ has a degree larger than the degree of any vertex not in~$R$.  Hence, on
input~$\widehat{G}$, the MDG algorithm first deletes all vertices from~$R$.
Subsequently, it can find a minimum vertex cover of~$H_2$, which has size
$\mvc(G_2) + \|E(G_2)\| (\maxdegree(G_2) + 1)$ by
Equation~(\ref{eq:vcsmdgone-2}), and eventually it can choose, say, the
vertices~$a_i$, $1 \leq i \leq j$, to cover the edges of~$B$.  Hence,
\begin{eqnarray*}
\minmdg(\widehat{G}) & = & 
 \|V(G_1)\| + \mvc(G_2) + \| E(G_2)\| (\maxdegree(G_2) + 1) + j \\
 & \stackrel{(\ref{eq:size-of-H2})}{=} & \mvc(G_2) + \| V(H_2)\| + j.
\end{eqnarray*}
We now consider $\mvc(\widehat{G})$.  
Since every vertex cover of $\widehat{G}$ must
contain all vertices of $L$ or all vertices of $R$ to cover the edges
connecting $L$ and~$R$, it follows from Equations~(\ref{eq:vcsmdgone-2})
and~(\ref{eq:size-of-H2}) that:
\begin{eqnarray*}
\mvc(\widehat{G})
 & = & \min\{ \|V(G_1)\| + \mvc(H_2) + j,\ \|V(H_2)\| + j + \mvc(G_1) \}  \\
 & = & \min\{ \mvc(G_2) + \|V(H_2)\| + j,\ \mvc(G_1) + \| V(H_2)\| + j\}.
\end{eqnarray*}
Since $\mvc(G_2) \geq \mvc(G_1)$, it follows that
\[
        \mvc(\widehat{G}) = \mvc(G_1) + \| V(H_2)\| + j.
\]
Hence, setting $q = \| V(H_2)\| + j$, Equations~(\ref{eq:minmdg-special})
and~(\ref{eq:mvcmdg-special}) are satisfied, which completes the proof that
$\smdgone$ is $\parallelnp$-complete.

We now turn to the proof that $\smdgr$ is $\parallelnp$-complete for $r>1$.
Fix any rational number $r = \frac{\ell}{m}$, where $\ell$ and $m$ are
integers with $1 \leq m < \ell$.  Without loss of generality, we may assume
that $\gcd(\ell - m, m) = 1$, where $\gcd(a, b)$ denotes the greatest common
divisor of the integers $a$ and~$b$.
Recall that the pair $\pair{G_1, G_2} = f(x)$ of graphs is obtained using the
reduction $f$ from $X$ to $\vcgeq$ according to Lemma~\ref{lem:vcgeq}; hence,
$\mvc(G_2) \geq \mvc(G_1)$.  

We will define a graph $\widehat{G}_{r}$ and integers $p, q \geq 0$ such that:
\begin{eqnarray}
\label{eq:G1G2-minmdg}
\minmdg(\widehat{G}_{r}) & = & r (p \cdot \mvc(G_2) + q); \\
\label{eq:G1G2-mvc}
\mvc(\widehat{G}_{r})   & = & p \cdot \mvc(G_1) + q.
\end{eqnarray}

The reduction mapping any given string $x$ (via the pair $\pair{G_1, G_2}$
obtained according to Lemma~\ref{lem:vcgeq}) to the graph 
$\widehat{G}_{r}$ such that
Equations~(\ref{eq:G1G2-minmdg}) and~(\ref{eq:G1G2-mvc}) are satisfied will
establish that $X \manyone \smdgr$.  In particular, from these equations, we
have that:
\begin{itemize}
\item $\mvc(G_2) = \mvc(G_1)$ implies $\minmdg(\widehat{G}_{r}) = r \cdot
  \mvc(\widehat{G}_{r})$, and
\item $\mvc(G_2) > \mvc(G_1)$ implies $\minmdg(\widehat{G}_{r}) > r \cdot
  \mvc(\widehat{G}_{r})$.
\end{itemize}

We now describe the construction of~$\widehat{G}_{r}$:
\begin{itemize}
\item
  Let $g$ be the reduction from
  Lemma~\ref{lem:vcsmdgone} and 
  let $H_2 = g(G_2)$.
  Thus, $H_2 \in \smdgone$ and Equation~(\ref{eq:vcsmdgone-2}) holds:
\[
\mvc(H_2) = \mvc(G_2) + \| E(G_2) \| (\maxdegree(G_2) + 1).
\]

\item Let $G_{1}^{1}, G_{1}^{2}, \ldots, G_{1}^{m}$ be $m$ pairwise disjoint
  copies of~$G_1$, and let $H_{2}^{1}, H_{2}^{2}, \ldots, H_{2}^{\ell}$ be
  $\ell$ pairwise disjoint copies of~$H_2$.
  
\item Let $\tilde{U} = \bigcup_{i = 1}^{\ell} H_{2}^{i}$ be the disjoint union
  of these copies of~$H_2$, and rename the vertices of $\tilde{U}$ by
  $V(\tilde{U}) = \{ \tilde{u}_1, \tilde{u}_2, \ldots ,
  \tilde{u}_{\ell \cdot \|V(H_2)\|} \}$.
  
\item Let $Z = \bigcup_{i = 1}^{m} G_{1}^{i}$ be the disjoint union of
  these copies of~$G_1$, and rename the vertices of $Z$ by $V(Z) = \{ z_1,
  z_2, \ldots , z_{m \cdot \|V(G_1)\|} \}$.
  
\item To apply Lemma~\ref{lem:mdg-bipartite}, choose $n_1 = \ell \cdot
  \|V(H_2)\|$, $n_2 \geq m \cdot \|V(G_1)\|$, and $\delta = \maxdegree(H_2) +
  1$, where the exact value of $n_2$ will be specified below.  Choose the
  constant $\mu$ so as to satisfy Equation~(\ref{eq:mdg-bipartite}):
\[
\mu (\ln \mu - 2 \ln (\delta + 2) - 1) \geq n_2 + n_1 .
\]
    
\item Given the constants $n_1$, $n_2$, $\delta$, and~$\mu$, 
  define $\widehat{G}_{r}$
  to be the bipartite graph $G$ from Lemma~\ref{lem:mdg-bipartite} extended by
  the edges between the $\tilde{u}_i$ vertices that were added above to
  represent the structure of the copies of~$H_2$, and extended by the edges
  between the $z_j$ vertices that were added above to represent the structure
  of the copies of~$G_1$.  
  That is, unlike~$G$, the graph $\widehat{G}_{r}$ is no
  longer a bipartite graph.  
  Formally, the vertex set of $\widehat{G}_{r}$ is given by
\begin{eqnarray*}
V(\widehat{G}_{r}) & = & V(G) = V \cup \tilde{V} , \ \ \mbox{ where} \\
V & = &  \{ u_1, u_2, \ldots , u_{n_1}, w_1, w_2, \ldots , w_{\mu}, z_1,
z_2, \ldots z_{n_2}\} \ \ \mbox{ and} \\
\tilde{V} & = & \{ \tilde{u}_1, \tilde{u}_2,
\ldots , \tilde{u}_{n_1}, \tilde{w}_1, \tilde{w}_2, \ldots ,
\tilde{w}_{\mu} \} ,\end{eqnarray*}
and the edge set of $\widehat{G}_{r}$ is given by
$E(\widehat{G}_{r}) = E(G) \cup E(\tilde{U}) \cup E(Z)$,
where $E(G)$ is constructed as in the proof of Lemma~\ref{lem:mdg-bipartite}.
\end{itemize}

This completes the construction of~$\widehat{G}_{r}$.  We now prove
Equations~(\ref{eq:G1G2-minmdg}) and~(\ref{eq:G1G2-mvc}).

\begin{enumerate}
\item \label{enum:smdgr-1} We first consider $\minmdg(\widehat{G}_{r})$.  By
  construction, for each vertex $v$ in $\tilde{V}$, we have
\begin{equation}
\label{eq:prop4}
\degree_{\widehat{G}_{r}}(v) \leq 
\degree_{G}(v) + \maxdegree(H_2) < \degree_{G}(v) +
\delta.
\end{equation}
Let $S$ be any induced subgraph of $\widehat{G}_{r}$ 
that can be obtained by deleting
vertices from~$V$ such that $V \cap V(S) \neq \emptyset$.
Property~\ref{eq:mdg-bipartite-4} of Lemma~\ref{lem:mdg-bipartite} and
Equation~(\ref{eq:prop4}) imply that
\[
\max_{v \in V \cap V(S)} \degree_S(v) > \max_{v \in \tilde{V}} \degree_S(v).
\]
Hence, on input~$\widehat{G}_{r}$, 
the MDG algorithm starts by choosing the $n_1 + \mu + n_2$ 
vertices from~$V$, which isolates each vertex $\tilde{w}_i \in
\tilde{V}$ and leaves $\ell$ isolated copies of $H_2$.  Subsequently, since
$H_2 \in \smdgone$, the MDG algorithm can choose a minimum vertex cover in
each of these $\ell$ copies of $H_2$.  By Equation~(\ref{eq:vcsmdgone-2}),
\[
\mvc(H_2) = \mvc(G_2) + \| E(G_2) \| (\maxdegree(G_2) + 1) ,
\]
and hence,
\[
\minmdg(\widehat{G}_{r}) = n_1 + \mu + n_2 + \ell(\mvc(G_2) + \|E(G_2)\|
(\maxdegree(G_2) + 1)) .
\]

\item \label{enum:smdgr-2} We now consider $\mvc(\widehat{G}_{r})$. 
   Define the set $C = \tilde{V} \cup D$, 
  where $D$ with $\|D\| = m \cdot \mvc(G_1)$ is a
  minimum vertex cover of~$Z$.  It is obvious from the construction of
  $\widehat{G}_{r}$ that $C$ is a minimum vertex cover 
  of~$\widehat{G}_{r}$.  Hence,
\[
\mvc(\widehat{G}_{r}) = n_1 + \mu + m \cdot \mvc(G_1).
\]
\end{enumerate}
   
To complete the proof, 
we have to choose $n_2 \geq m \cdot \|V(G_1)\|$ such that
Equations~(\ref{eq:G1G2-minmdg}) and~(\ref{eq:G1G2-mvc}) are satisfied for
suitable integers $p$ and~$q$.
Setting $p = m$ and $q = n_1 + \mu$ and requiring
\begin{equation}  \label{mdg::eq::gl3}
n_1 + n_2 + \mu + \ell \cdot \|E(G_2)\| (\maxdegree(G_2) + 1)
 = r(n_1 + \mu )
\end{equation}
or, equivalently,
\begin{equation} \label{mdg::eq::gl4}
        m \cdot n_2 + m \cdot \ell \cdot \|E(G_2)\| (\maxdegree(G_2) + 1))
 = (\ell - m) n_1 + (\ell - m) \mu
\end{equation}
satisfies Equations~(\ref{eq:G1G2-minmdg}) and~(\ref{eq:G1G2-mvc}).  Our
assumption that $\gcd(\ell - m, m) = 1$ implies that
Equation~(\ref{mdg::eq::gl4}) has integer solutions in the variables $n_2$
and~$\mu$.  It is easy
to see that one such solution, say $(n_2, \mu)$, simultaneously (a)~satisfies
Equation~(\ref{eq:mdg-bipartite}), (b)~satisfies that both $n_2$
and~$\mu$ are polynomially bounded in the size of the input of the reduction
being described, and (c)~can be computed 
efficiently~\cite{cla-for:j:efficient-solution-diophantine-equations}.
This completes the proof of the theorem.~\qed

\bigskip

\noindent {\bf Acknowledgments:}  
We thank Dieter Kratsch and Andreas Brandst\"adt for interesting discussions
on graph theory and graph-theoretical notation.

\bibliography{/home/inf1/rothe/BIGBIB/joergbib}

\end{document}

%% file: sedr.eepic
\setlength{\unitlength}{0.00045in}
\begingroup\makeatletter\ifx\SetFigFont\undefined%
\gdef\SetFigFont#1#2#3#4#5{%
  \reset@font\fontsize{#1}{#2pt}%
  \fontfamily{#3}\fontseries{#4}\fontshape{#5}%
  \selectfont}%
\fi\endgroup%
{\renewcommand{\dashlinestretch}{30}
\begin{picture}(10683,4056)(0,-10)
\thicklines
\put(5872,914){\blacken\ellipse{180}{180}}
\put(5872,914){\ellipse{180}{180}}
\put(5872,104){\blacken\ellipse{180}{180}}
\put(5872,104){\ellipse{180}{180}}
\path(5872,1724)(5872,104)
\path(5872,1724)(5872,104)
\put(6547,914){\blacken\ellipse{180}{180}}
\put(6547,914){\ellipse{180}{180}}
\put(6547,104){\blacken\ellipse{180}{180}}
\put(6547,104){\ellipse{180}{180}}
\path(6547,1724)(6547,104)
\path(6547,1724)(6547,104)
\put(9472,914){\blacken\ellipse{180}{180}}
\put(9472,914){\ellipse{180}{180}}
\put(9472,104){\blacken\ellipse{180}{180}}
\put(9472,104){\ellipse{180}{180}}
\path(9472,1724)(9472,104)
\path(9472,1724)(9472,104)
\put(697,3344){\ellipse{900}{900}}
\put(2047,3344){\ellipse{900}{900}}
\put(3847,3344){\ellipse{900}{900}}
\put(6097,3344){\ellipse{900}{900}}
\put(7447,3344){\ellipse{900}{900}}
\put(9247,3344){\ellipse{900}{900}}
\put(2272,1994){\ellipse{2250}{900}}
\put(7537,2129){\ellipse{2250}{900}}
\put(127,1424){\arc{210}{1.5708}{3.1416}}
\put(127,3914){\arc{210}{3.1416}{4.7124}}
\put(4417,3914){\arc{210}{4.7124}{6.2832}}
\put(4417,1424){\arc{210}{0}{1.5708}}
\path(22,1424)(22,3914)
\path(127,4019)(4417,4019)
\path(4522,3914)(4522,1424)
\path(4417,1319)(127,1319)
\put(5527,1424){\arc{210}{1.5708}{3.1416}}
\put(5527,3914){\arc{210}{3.1416}{4.7124}}
\put(9817,3914){\arc{210}{4.7124}{6.2832}}
\put(9817,1424){\arc{210}{0}{1.5708}}
\path(5422,1424)(5422,3914)
\path(5527,4019)(9817,4019)
\path(9922,3914)(9922,1424)
\path(9817,1319)(5527,1319)
\put(407,3254){\makebox(0,0)[lb]{\smash{{{\SetFigFont{12}{14.4}{\rmdefault}{\mddefault}{\updefault}$H_2^1$}}}}}
\put(1757,3254){\makebox(0,0)[lb]{\smash{{{\SetFigFont{12}{14.4}{\rmdefault}{\mddefault}{\updefault}$H_2^2$}}}}}
\put(7157,3254){\makebox(0,0)[lb]{\smash{{{\SetFigFont{12}{14.4}{\rmdefault}{\mddefault}{\updefault}$H_1^2$}}}}}
\put(5807,3254){\makebox(0,0)[lb]{\smash{{{\SetFigFont{12}{14.4}{\rmdefault}{\mddefault}{\updefault}$H_1^1$}}}}}
\put(2737,3299){\makebox(0,0)[lb]{\smash{{{\SetFigFont{12}{14.4}{\rmdefault}{\mddefault}{\updefault}$\cdots$}}}}}
\put(8122,3284){\makebox(0,0)[lb]{\smash{{{\SetFigFont{12}{14.4}{\rmdefault}{\mddefault}{\updefault}$\cdots$}}}}}
\put(4237,1514){\makebox(0,0)[lb]{\smash{{{\SetFigFont{12}{14.4}{\rmdefault}{\mddefault}{\updefault}$L$}}}}}
\put(5512,1499){\makebox(0,0)[lb]{\smash{{{\SetFigFont{12}{14.4}{\rmdefault}{\mddefault}{\updefault}$R$}}}}}
\put(7447,644){\makebox(0,0)[lb]{\smash{{{\SetFigFont{12}{14.4}{\rmdefault}{\mddefault}{\updefault}$\cdots$}}}}}
\put(1327,1904){\makebox(0,0)[lb]{\smash{{{\SetFigFont{12}{14.4}{\rmdefault}{\mddefault}{\updefault}indep.
set $I_2$}}}}}
\put(6622,2084){\makebox(0,0)[lb]{\smash{{{\SetFigFont{12}{14.4}{\rmdefault}{\mddefault}{\updefault}indep.
set $I_1$}}}}}
\put(4827,2669){\makebox(0,0)[lb]{\smash{{{\SetFigFont{12}{14.4}{\rmdefault}{\mddefault}{\updefault}$\bowtie$}}}}}
\put(3557,3254){\makebox(0,0)[lb]{\smash{{{\SetFigFont{12}{14.4}{\rmdefault}{\mddefault}{\updefault}$H_2^{\ell}$}}}}}
\put(8957,3254){\makebox(0,0)[lb]{\smash{{{\SetFigFont{12}{14.4}{\rmdefault}{\mddefault}{\updefault}$H_1^m$}}}}}
\put(6007,824){\makebox(0,0)[lb]{\smash{{{\SetFigFont{12}{14.4}{\rmdefault}{\mddefault}{\updefault}$a_1$}}}}}
\put(6682,824){\makebox(0,0)[lb]{\smash{{{\SetFigFont{12}{14.4}{\rmdefault}{\mddefault}{\updefault}$a_2$}}}}}
\put(9607,824){\makebox(0,0)[lb]{\smash{{{\SetFigFont{12}{14.4}{\rmdefault}{\mddefault}{\updefault}$a_{k\cdot\ell}$}}}}}
\put(6007,14){\makebox(0,0)[lb]{\smash{{{\SetFigFont{12}{14.4}{\rmdefault}{\mddefault}{\updefault}$b_1$}}}}}
\put(6682,14){\makebox(0,0)[lb]{\smash{{{\SetFigFont{12}{14.4}{\rmdefault}{\mddefault}{\updefault}$b_2$}}}}}
\put(9607,14){\makebox(0,0)[lb]{\smash{{{\SetFigFont{12}{14.4}{\rmdefault}{\mddefault}{\updefault}$b_{k\cdot\ell
}$}}}}}
\end{picture}
}

%% file: smdgone.eepic
\setlength{\unitlength}{0.00058333in}
\begingroup\makeatletter\ifx\SetFigFont\undefined%
\gdef\SetFigFont#1#2#3#4#5{%
  \reset@font\fontsize{#1}{#2pt}%
  \fontfamily{#3}\fontseries{#4}\fontshape{#5}%
  \selectfont}%
\fi\endgroup%
{\renewcommand{\dashlinestretch}{30}
\begin{picture}(7244,2926)(0,-10)
\thicklines
\put(1597,2214){\ellipse{900}{900}}
\put(697,1089){\blacken\ellipse{180}{180}}
\put(697,1089){\ellipse{180}{180}}
\put(1372,1089){\blacken\ellipse{180}{180}}
\put(1372,1089){\ellipse{180}{180}}
\put(2497,1089){\blacken\ellipse{180}{180}}
\put(2497,1089){\ellipse{180}{180}}
\put(697,189){\blacken\ellipse{180}{180}}
\put(697,189){\ellipse{180}{180}}
\put(1372,189){\blacken\ellipse{180}{180}}
\put(1372,189){\ellipse{180}{180}}
\put(2497,189){\blacken\ellipse{180}{180}}
\put(2497,189){\ellipse{180}{180}}
\put(127,744){\arc{210}{1.5708}{3.1416}}
\put(127,2784){\arc{210}{3.1416}{4.7124}}
\put(3067,2784){\arc{210}{4.7124}{6.2832}}
\put(3067,744){\arc{210}{0}{1.5708}}
\path(22,744)(22,2784)
\path(127,2889)(3067,2889)
\path(3172,2784)(3172,744)
\path(3067,639)(127,639)
\put(4177,744){\arc{210}{1.5708}{3.1416}}
\put(4177,2784){\arc{210}{3.1416}{4.7124}}
\put(7117,2784){\arc{210}{4.7124}{6.2832}}
\put(7117,744){\arc{210}{0}{1.5708}}
\path(4072,744)(4072,2784)
\path(4177,2889)(7117,2889)
\path(7222,2784)(7222,744)
\path(7117,639)(4177,639)
\path(697,1044)(697,189)
\path(697,1044)(697,189)
\path(1372,1044)(1372,144)
\path(1372,1044)(1372,144)
\path(2497,1044)(2497,144)
\path(2497,1044)(2497,144)
\put(3577,1989){\makebox(0,0)[lb]{\smash{{{\SetFigFont{12}{14.4}{\rmdefault}{\mddefault}{\updefault}$\!\!\bowtie$}}}}}
\put(1507,2124){\makebox(0,0)[lb]{\smash{{{\SetFigFont{12}{14.4}{\rmdefault}{\mddefault}{\updefault}$\!\!H_2$}}}}}
\put(607,1314){\makebox(0,0)[lb]{\smash{{{\SetFigFont{12}{14.4}{\rmdefault}{\mddefault}{\updefault}$a_1$}}}}}
\put(1282,1314){\makebox(0,0)[lb]{\smash{{{\SetFigFont{12}{14.4}{\rmdefault}{\mddefault}{\updefault}$a_2$}}}}}
\put(1750,344){\makebox(0,0)[lb]{\smash{{{\SetFigFont{12}{14.4}{\rmdefault}{\mddefault}{\updefault}$\cdots$}}}}}
\put(2407,1314){\makebox(0,0)[lb]{\smash{{{\SetFigFont{12}{14.4}{\rmdefault}{\mddefault}{\updefault}$a_j$}}}}}
\put(382,54){\makebox(0,0)[lb]{\smash{{{\SetFigFont{12}{14.4}{\rmdefault}{\mddefault}{\updefault}$b_1$}}}}}
\put(1057,54){\makebox(0,0)[lb]{\smash{{{\SetFigFont{12}{14.4}{\rmdefault}{\mddefault}{\updefault}$b_2$}}}}}
\put(2600,54){\makebox(0,0)[lb]{\smash{{{\SetFigFont{12}{14.4}{\rmdefault}{\mddefault}{\updefault}$b_j$}}}}}
\put(2587,1989){\makebox(0,0)[lb]{\smash{{{\SetFigFont{12}{14.4}{\rmdefault}{\mddefault}{\updefault}$L$}}}}}
\put(4567,1989){\makebox(0,0)[lb]{\smash{{{\SetFigFont{12}{14.4}{\rmdefault}{\mddefault}{\updefault}$R=G_1$}}}}}
\end{picture}
}